\documentclass{article}

\usepackage[english]{babel}
\usepackage[utf8x]{inputenc}
\usepackage[T1]{fontenc}

\usepackage[a4paper,top=3cm,bottom=2cm,left=3cm,right=3cm,marginparwidth=1.75cm]{geometry}

\usepackage{amsmath}
\usepackage{graphicx}
\usepackage[colorlinks=true, allcolors=blue]{hyperref}
\usepackage{amssymb}

\usepackage{comment}
\usepackage[natbibapa]{apacite}
\usepackage{subcaption}
\usepackage{setspace}
\usepackage{bibentry}
\usepackage{authblk}

\title{Effects of spatial smoothing on group-level differences in functional brain networks}

\author[1,*]{Ana Mar{\'i}a Triana}
\author[2]{Enrico Glerean}
\author[1]{Jari Saram{\"a}ki}
\author[3]{Onerva Korhonen}
\affil[1]{Department of Computer Science, School of Science, Aalto University, Espoo, Finland}
\affil[2]{Department of Neuroscience and Biomedical Engineering, School of Science, Aalto University, Espoo, Finland}
\affil[3]{Universit{\'e} de Lille, CNRS, UMR 9193 - SCALab - Sciences Cognitives et Sciences Affectives, Lille, France}
\affil[*]{Corresponding author: Department of Computer Science, School of Science, Aalto University, Espoo, Finland. \textit{E-mail address:} ana.trianahoyos@aalto.fi (A.M. Triana)}
\date{}                     %% if you don't need date to appear
\begin{document}

\maketitle

\begin{abstract}

Brain connectivity with functional Magnetic Resonance Imaging (fMRI) is a popular approach for detecting differences between healthy and clinical populations. Before creating a functional brain network, the fMRI time series must undergo several preprocessing steps to control for artifacts and to improve data quality. However, preprocessing may affect the results in an undesirable way. Spatial smoothing, for example, is known to alter functional network structure. Yet, its effects on group-level network differences remain unknown.

Here, we investigate the effects of spatial smoothing on the difference between patients and controls for two clinical conditions: autism spectrum disorder and bipolar disorder, considering fMRI data smoothed with Gaussian kernels (0-32 mm).

We find that smoothing affects network differences between groups. For weighted networks, incrementing the smoothing kernel makes networks more different. For thresholded networks, larger smoothing kernels lead to more similar networks, although this depends on the network density. Smoothing also alters the effect sizes of the individual link differences. This is independent of the ROI size, but vary with link length.

The effects of spatial smoothing are diverse, non-trivial, and difficult to predict. This has important consequences: the choice of smoothing kernel affects the observed network differences.

\end{abstract}

\section{Introduction}

Neuroscientists commonly model the brain as a network. The brain can be seen as a system of segregated, specialized neuronal groups that interact to produce intricate behaviors \citep{SPORNS2013162}. These groups and their interactions form a network that can be studied with the tools of network science. One of the applications of brain network studies is to investigate connectivity similarities between subjects and groups \citep{Bullmore2009}. This may ultimately lead to the identification of abnormalities in the connectivity of clinical populations \citep{Bassett2009}.

In the network approach, brain regions are used as nodes and links are defined as structural or functional connections \citep{Spoorns}. Since nodes and links can be defined in different ways \citep{Stanley2013,Bullmore2009}, the network approach can be applied to different imaging techniques, making it a versatile tool for analyzing brain functions. For example, functional magnetic resonance imaging (fMRI) has been repeatedly adopted to analyze brain network patterns \citep{VANDENHEUVEL2010519,Bullmore2009}.

fMRI relies on the Blood-oxygen-level-dependent (BOLD) signal, which is an indirect measure of neural activity \citep{Kwong1992,Ogawa1992}. Unfortunately, fMRI features numerous sources of undesired variability such as head motion, respiratory and cardiac cycles, thermal noise, and hardware artifacts \citep{Murphy2013} that cause noise in the BOLD signal. Consequently, fMRI preprocessing steps are crucial for cleaning the signal before constructing any brain network. However, choosing the optimal preprocessing pipeline from the myriad of possibilities is difficult \citep{doi:10.1146/annurev-clinpsy-040510-143934}. Several preprocessing steps can directly affect the results. For example, \cite{doi:10.1089/brain.2014.0292} show that registration approaches have a strong impact on network formation, and \cite{Power2012} and \cite{POWER2014320} highlight the importance of head motion correction since head motion can alter functional connectivity. Further, \cite{MURPHY2017169} review the impacts of global signal regression in resting-state fMRI (rsfMRI), and \cite{10.3389/fncom.2018.00008} report that graph-theoretical measures of functional connectivity depend on the order and choice of preprocessing steps. Despite these findings, few researchers comment on the possible effects of their preprocessing choices. In particular, little research on the topic has been translated to the clinical setting.

One of the commonly used preprocessing steps involves artificially increasing the spatial smoothness of the data. We will refer to this preprocessing step as \emph{spatial smoothing}. In spatial smoothing, the signal of each voxel is averaged with the signal of its neighbors, weighted by a Gaussian kernel of some chosen full width at half maximum (FWHM). Spatial smoothing aims to compensate for inaccuracies in spatial registration, increase the signal-to-noise ratio (SNR), and decrease inter-subject variability when the analysis paradigm is the general linear model (GLM) \citep{Pajula2018, Mikl2018, HOPFINGER2000326, doi:10.1111/j.1749-6632.2010.05446.x}. Note that independently of this preprocessing step, all fMRI data have some intrinsic level of spatial smoothness, determined by the point spread function of the measuring scanner~\citep{FRIEDMAN20061656}. Therefore, the spatial smoothness of data is always a combination of this intrinsic smoothness and the additional smoothness induced by spatial smoothing in preprocessing.

Different levels of spatial smoothness are known to affect functional network properties \citep{Stanley2013, FORNITO2013426}. Increasing the level of smoothness with spatial smoothing in preprocessing may increase the similarity between voxel time courses and suppress the fluctuation amplitudes in seed connectivity analysis \citep{doi:10.1089/brain.2011.0018}. Spatial smoothing also affects the degrees, the distribution of link lengths, and the composition of the largest connected component \citep{doi:10.1111/ejn.13717}. Because spatial smoothing is known to affect the properties of individual functional networks, it is important to know whether it affects group-level differences too.

To answer this question, we use rsfMRI data from 94 male subjects from the Autism Brain Imaging Data Exchange (ABIDE) database \citep{DiMartino2017,DiMartino2014}. The subjects are divided into two groups (N=47, age-matched pairs): typical controls (TC) and Autism Spectrum Disorder patients (ASD). fMR images are smoothed using Gaussian kernels with FWHM from 0 mm to 32 mm. We assess group differences using Network Based Statistics (NBS) \citep{ZALESKY20101197} and compare the results obtained for all spatial smoothing levels. In addition, to address the generalizability of the results, we repeat the analysis for fMRI data from 44 subjects from the UCLA Consortium for Neuropsychiatric Phenomics LA5c Study \citep{Gorgolewski2017}. These subjects are also divided into two groups (N=22): bipolar disorder patients and typical controls.

For full, weighted networks (correlation matrices without thresholding), we find that the difference in link structure between groups increases with the FWHM. Only few links are significantly different at all smoothing levels where group differences are found. Moreover, the effect size of the differences varies with the kernel size. Surprisingly, the effects of spatial smoothing are independent of ROI size. On the other hand, the effects of smoothing vary depending on link length, although irregularly. Spatial smoothing affects group comparison results in thresholded, weighted networks too, although these effects depend highly on network density.

Our results show that spatial smoothing affects differences in network structure between subject groups. These effects are non-trivial and diverse. Our results are pivotal if network analysis is used to find a \textit{network fingerprint} of a disease, as the undesired effects of spatial smoothing could lead to spurious results in these analyses.

\section{Materials and Methods}

\subsection{Data}
\label{sec:data}

\subsubsection*{Autism Brain Imaging Data Exchange (ABIDE).}
To investigate the effects of spatial smoothing on group-level differences in functional brain networks, we employ unpreprocessed data from the Autism Brain Imaging Data Exchange (ABIDE) \citep{DiMartino2014, DiMartino2017}. This dataset contains anonymized MRI and rsfMRI images and phenotypic information from 2156 individuals, collected at 19 different institutions. The subjects are classified as typical healthy controls (TC) or diagnosed with the Autism Spectrum Disorder (ASD). ABIDE provides several advantages. First, the use of open data makes the replication of the study easier. Second, using data from different institutions helps to take possible multisite effects into account. Finally, since ABIDE has been used in many analyses to identify differences between ASD and TC populations, it supplies a baseline for comparison \citep{Maximo2014}. The sites and the voxel size of their images is listed in Supplementary Table 1.

Subjects were selected based on three initial criteria. (i) In terms of gender, we chose only male subjects due to the prevalence of ASD in males \citep{Faras2010}; (ii) Regarding age, we excluded subjects below 18 years, since brain networks evolve with age, in particular in childhood and adolescence \citep{VASA2016245}; and finally, (iii) in terms of repetition time (TR), we chose data acquired with TR=2 s to reduce the analysis complexity. These initial limitations yielded 231 potential subjects, whose images were preprocessed using BRAMILA v2.0 (see below) \citep{Bramila}.

As an Image Quality Control measure, we visually inspected the images before preprocessing. We discarded 50 subjects whose images had noticeable artifacts, like high motion, missing brain regions, or regions with unexplained changes in intensity. To assure data quality, we excluded subjects whose images contained noticeable head motion, since it can alter the rsfMRI analysis \citep{Power2012} (examples of these images are found in Supplementary Figure A1). Detailed explanations about the image quality control can be found in the Supplementary Information, section Image quality control.

After preprocessing (see below), we checked the MCFLIRT results for any peaks over $\pm$ 2 mm and over 0.04 radians. However, no images were excluded because of this criteria. We discarded subjects whose rsfMRI sequence did not contain a minimum of continuous 4.5 minutes without large framewise displacement peaks (FD$>0.5$). The procedure for calculating the framewise displacement followed that of \cite{Power2012}. However, the applied inclusion criterion was different: we considered only continuous temporal intervals in which the framewise displacement has not exceed 0.5 according to the temporal masks. This limitation yielded 128 subjects. Then, we matched the ages of the remaining subjects, so that the age difference between patient-control pairs is no greater than 9 years, avoiding mixing between measurement institutions (e.g. ASD subject from ETH Z\"urich is not matched with a TC subject from NYU Langone Medical Center) \citep{DiMartino2017}.
Finally, following \cite{DANSEREAU2017220}, we verified that our final sample had as many subjects as possible with balanced numbers between groups. These additional limitations yielded 94 male subjects (47 subjects diagnosed with ASD, and 47 healthy controls), age 24.15$\pm$5.58 years. The complete list of subjects (discarded, preprocessed, and age-matched) and the reasons for exclusion are listed in the Supplementary Table 2.

\subsubsection*{UCLA Consortium for Neuropsychiatric Phenomics LA5c Study.}

To verify that our results generalize to other datasets, we repeated all analysis using unpreprocessed data from a second, independent dataset. This dataset, UCLA, contains fMRIs  of 272 subjects divided in 4 populations: healthy controls (130 subjects), patients diagnosed with ADHD (43 subjects), bipolar disorder (49 subjects), and schizophrenia (50 subjetcs) \citep{Gorgolewski2017}. We chose to compare patients diagnosed with bipolar disorder and healthy controls, as the bipolar population had the largest number of scans that comply with our restrictions in image quality control and framewise displacement. Out of the 272 initial subjects, 8 were discarded for missing T1-weighted or rsfMRI images, 93 were discarded because they were not bipolar or controls, and 80 were not selected for not complying with the framewise displacement requirement. After matching the subjects in gender and age, the sample has 22 patients diagnosed with bipolar disorder and 22 healthy controls. The list of subject IDs and reasons for inclusion/exclusion can be found in Supplementary Table 3.

\subsection{Data preprocessing}

The data preprocessing has two parts (see Fig.~\ref{fig:methods} A): image preprocessing and spatial smoothing. In the following sections, we describe each preprocessing step in detail.  

\subsubsection{Image preprocessing.}
\label{sec:bramila}
ABIDE Structural MRI data were preprocessed with the FSL software (\url{www.fmrib.ox.ac.uk}, version 5.0.9) \citep{JENKINSON2012782, WOOLRICH2009S173, SMITH2004S208}. The T1-weighted images were segmented into gray matter (GM), white matter (WM), and cerebrospinal fluid (CSF), whilst also correcting for radiofrequency field inhomogeneities using the FMRIB Automated Segmentation Tool (FAST) \citep{FAST}. Then, non-brain tissue was deleted from the image with the FSL brain extraction tool (BET) \citep{HBM:HBM10062}. UCLA Structural MRI data were preprocessed using the fmriprep T1w preprocessing workflow \citep{esteban_oscar_2017_996169}, which apply the following steps: brain extraction, brain tissue segmentation, and registration to the MNI space.

fMRI data were preprocessed using the FSL software and the BRAMILA pipeline. First, EPI slices were corrected for slice timing differences according to each institution's acquisition sequence. Then, volumes were corrected for head motion by means of MCFLIRT \citep{JENKINSON2002825}. Afterward, the images were co-registered to the Montreal Neurological Institute 152 2 mm template in a two-step registration procedure using FLIRT: from EPI to the subject anatomical image after brain extraction (9 degrees of freedom) and from anatomical to the standard template (12 degrees of freedom) \citep{JENKINSON2002825, Jenkinson2018}. A 240-s-long Savitzky-Golay filter \citep{Cukur2013} was applied to remove scanner drift. Further, the BOLD time series were cleaned using 24 motion-related regressors, signal from deep WM, ventricles and CSF locations to control for motion and physiological artifacts, following \cite{Power2012}. Later, the time series were filtered with a Butterworth filter (0.01-0.08 Hz). Finally, spatial smoothing was applied with 16 different Gaussian kernels.

\subsubsection{Smoothing.}
Spatial smoothing aims to increase the SNR and it is usually the last implemented preprocessing step. In spatial smoothing, each voxel signal is redefined as the average of the signals from the voxel and its neighbors, weighted by a smoothing kernel:
\begin{equation}
x_{i}(t)=\frac{\sum_{j}{G_{i}(j)x_{j}(t)}}{\sum_{j}{G_{i}(j)}},
\label{eq:smoothing}
\end{equation}
where $x_{i}(t)$ is the time series of voxel $i$, $G_{i}(j)$ is the value of the smoothing kernel $G_{i}$ centered at voxel $i$ evaluated at voxel $j$, $x_{j}(t)$ denotes the time series of voxel $j$, and the summation is over all voxels. For most of the voxels $j$, $G_{i}(j)\approx0$. The distance (in mm) at which the filter operates is expressed by the FWHM. We used 16 different Gaussian kernels from 0 to 32 mm to investigate the impact of spatial smoothing in network comparison between groups. We included rather large smoothing kernels to provide a wider perspective, even though  most fMRI studies use only one kernel, whose FWHM typically varies from 4 to 12 mm.

This commonly used method smooths the data independently of its intrinsic smoothness, which can vary between measurement sites and subjects. Because of this, we also separately employed AFNI 3dBlurToFWHM \citep{AFNI} to take into account the influence of the intrinsic smoothness of the data. This function smooths an image until it reaches a specified smoothness. At each iteration, the smoothness of the data is estimated globally. Voxels where the local smoothness is higher than the goal will not be further smoothed. As the global smoothness approximates to the target value, the smoothing rate is decreased until the target smoothness is reached. This method is a form of adaptive smoothing in which each image in a dataset is smoothed with a different kernel, so that the smoothness variation within the dataset is decreased.

\begin{figure}[!h]
\center{\includegraphics[width=\textwidth]{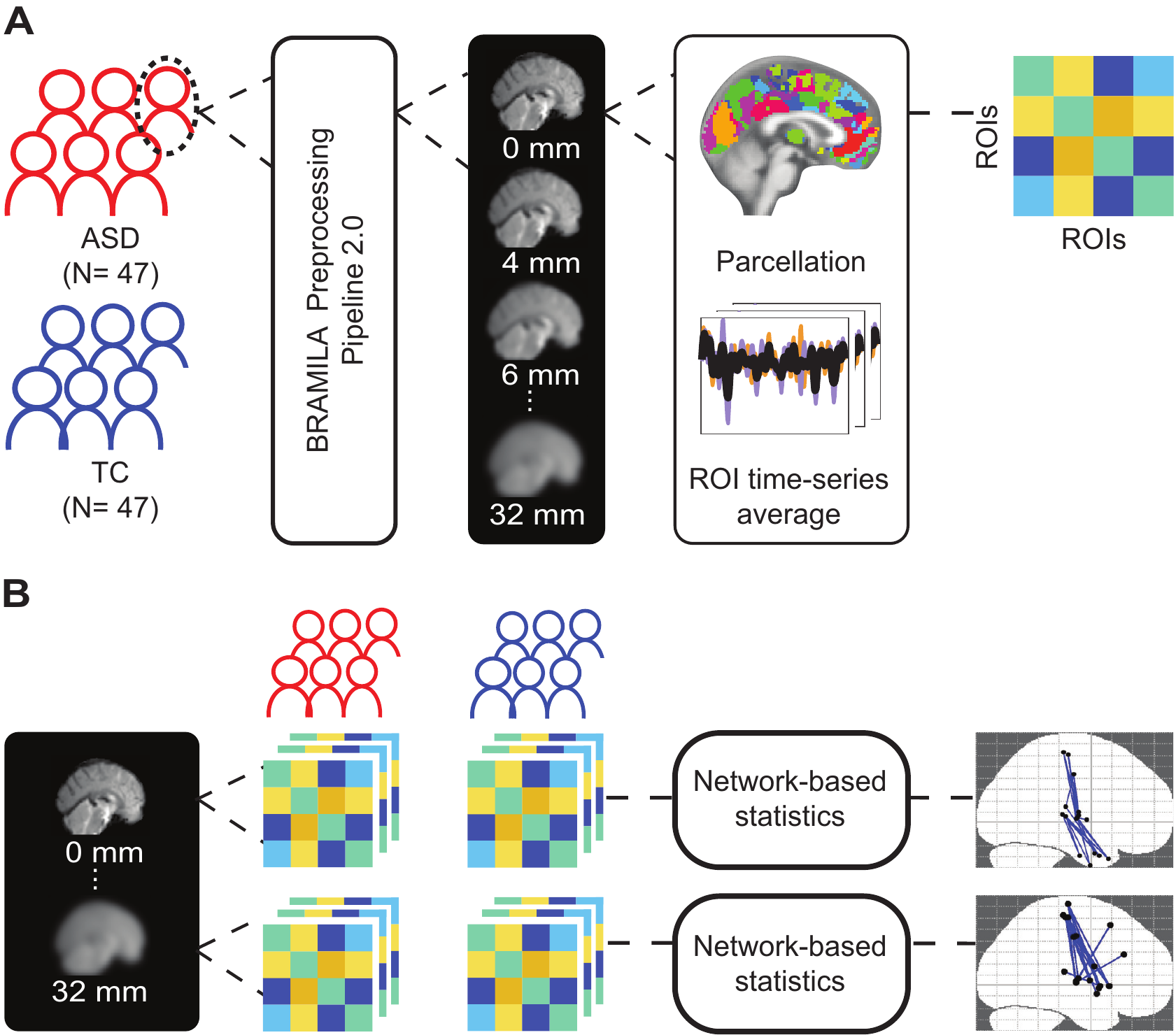}}
\caption{\label{fig:methods} Outline of data preprocessing and detection of network differences. (A) Subjects are divided in two groups according to their diagnosis: Autism Spectrum Disorder (ASD) or Typical Controls (TC). For each subject, functional networks are constructed by applying a series of steps. First, the fMRI data is preprocessed using a standard preprocessing pipeline (\citep{POWER2014320} and FSL) without applying spatial smoothing (FWHM=0 mm). Next, the images are smoothed with 16 different kernels, starting from 4 mm and increasing in 2 mm steps; data with no additional smoothing is also included (FWHM=0 mm). Then, for each of these images, we average the voxel time-series according to the Regions of Interest (ROI). Finally, we calculate the connectivity matrix as Pearson correlations between the averaged ROI time-series. This process yields one connectivity matrix per subject per smoothing kernel. (B) The matrices are organized according to the subjects' diagnosis for each smoothing kernel. This grouping creates 16 models that are individually fed to the Network-based Statistics toolbox (NBS) \citep{ZALESKY20101197} to investigate the group-level differences in functional networks for each level of spatial smoothing.}
\end{figure}

\subsection{Network analysis}
\label{sec:network}
\subsubsection{Network extraction.}
\label{sec:netextraction}
To compute the functional brain networks, we defined the nodes as the Regions of Interest (ROIs) from the Brainnetome atlas \citep{Fan2016}. The parcellation integrates different multimodal information, so structural and functional information are combined to provide a richer perspective of the human brain. This atlas comprises 210 cortical and 36 subcortical non-overlapping ROIs (i.e., each voxel was assigned to only one ROI), but it does not include cerebellar regions. In addition, we repeated our analysis with two different parcellations from the Craddock atlas. These parcellations are formed using the spatially-constrained normalized-cut spectral clustering algorithm \citep{Craddock2012}. Due to this algorithm, the parcellations may have less ROIs than the originally intended, for example, the parcellation K=100 contains only 98 ROIs. We chose parcellations with 98 (K=100) and 329 (K=350) ROIs as they were smaller and larger in number of ROIs than the Brainnetome Atlas (246 ROIs). We call these parcellations Craddock100 and Craddock350, respectively. Information about the parcellations (e.g. ROI labels and ROI sizes) can be found in the Supplementary Tables 4, 5, 6, and 7.

The weights of links connecting the nodes are defined as the Pearson correlation coefficients between the averaged time series of the voxels belonging to a ROI. This procedure yields individual weighted adjacency matrices of size $n_{\mathrm{ROIs}} \times n_\mathrm{ROIs}$.

After computing the link weights, we applied the Fisher transform to the adjacency matrices to stabilize the variance for all the Pearson correlation values. Later, we implemented a regression model considering the places of scan and the head movement (measured as mean FD) as regressors for each link in the network \citep{DANSEREAU2017220,Power2012} for ABIDE. For UCLA, the regression model considers gender and head movement (FD) only, as all the images come from the same site. Finally, we applied the Fisher inverse transformation to the regressed adjacency matrices to obtain the connectivity matrices.

\subsubsection{Network comparison.}
\label{sec:netcomparison}
To identify functional connectivity differences between groups, we used the Networks Based Statistic (NBS) approach, introduced by \citet{ZALESKY20101197} (see Fig.~\ref{fig:methods} B). With this method, we can identify significantly different links that form a connected structure instead of individual links. This is advantageous because the network structure is taken into account when selecting statistically significant differences, unlike by False Discovery Rate (FDR).

In brief, NBS determines the statistically different connectivity structures between groups in six steps. First, a test statistic is computed for each link in the connectivity matrix. 
Second, these test statistics are thresholded according to a limit defined by the user, forming a set of suprathreshold links. 
Third, the breadth-first search algorithm \citep{ahuja1993network} is used to identify any possible connected components in the set of suprathreshold links; after each component is identified, its number of links is stored. Fourth, the membership of the groups is permutated and the previous steps are repeated M times. Fifth, for each permutation, the maximal component size is determined and stored, yielding an empirical estimate of the null distribution of maximal component size. Finally, the observed maximal component size is compared with this null distribution and its p-value is estimated; if this p-value is smaller than the level of significance set by the user, the connected component and its comprised links are declared significant. This set of significant connected links is called a subnetwork.

While applying NBS, we do not make any assumptions about under/over connectivity. We compute the F-statistic for different values (suprathresholds). In the main findings, we report links whose F-statistic is greater than 12.25 (F-value $>$ 12.25) with a 0.05 level of significance ($\alpha<0.05$). We always picked the first subnetwork, if NBS produced several.

We also carried out a more traditional network analysis, given that NBS is not the only method employed to assess differences between groups. Commonly, thresholded brain networks are summarized by a univariate metric which is then used to assess networks differences by a simple tests (e.g. a t-test) \citep{24324431}. We thresholded the individual connectivity matrices at 7\% density after the network extraction step and binarize them. Then, we compute 5 graph measures for each subject -- node degree (the number of a node's neighbors), the clustering coefficient (the density of links between a node's neighbors), global and local efficiency (the amount of network efficiency to exchange information), and betweenness centrality (the fraction of shortest paths passing through the node). We run non-parametric permutations tests \citep{doi:10.1002/hbm.1058} with 10000 permutations per measure. We also perform permutation tests on thresholded, weighted matrices (7\% density) to assess link disparity between groups. Finally, we control the FDR by using the approach introduced by \citet{10.2307/2346101}.

\subsubsection{Network differences visualization.}

We use circos \citep{Krzywinski18062009} to visualize the network differences between groups. To make the data representation easier, we sort the ROIs in 13 systems. These systems are defined for a 264 ROI atlas in \citet{POWER2011665}. To assign the system labels, we compute the distance between the ROI centroids of the two atlases and select the label of the closest ROI. For the Brainnetome atlas, if two or more labels match one region, we choose the label that is closest to the same region in the other hemisphere. We base our plots in the implementation of \citet{VanHorn2012}. The relevant code to replicate the plots is available in the Zenodo repository (see section Data and software availability). 

\subsection{Other metrics}
\label{sec:others}

To better understand how far the effects of smoothing go and the role played by the smoothing kernel width, we used three additional approaches: fixed-size ROI  analysis, effect sizes of links, and Hamming distances between the statistically significant differences.

\subsubsection{Fixed-size ROI analysis.}
To understand if the effects of spatial smoothing depend on ROI size, we created a set of equally-sized ROIs. These fabricated ROIs are spheres centered at the ROI centroids. Their size is fixed to 7 voxels, which is the maximum volume possible without the spheres overlapping. In this analysis, the spheres replace the Atlases ROIs in the Network extraction step (see section Network extraction). The rest of the analysis has no alterations.

\subsubsection{Effect size.}
To quantify the magnitude of the differences between groups, we use Cohen's $d$ measure of effect size \citep{cohen88} computed for each significant link at all levels of smoothing. This measure can be calculated from the t-test statistics (t) as $d = t\sqrt{1/n_1 + 1/n_2}$, where $n_1$ and $n_2$ are the sample sizes of the group 1 and 2, respectively \citep{Lakens2013}. Nevertheless, given the degree of freedom of our experiment (df=1), we can assume the equivalency between F-test and T-test values $F=t^2$, so the formula becomes $d = \sqrt{F\big(\frac{1}{n_1} + \frac{1}{n_2}\big)}$.

\subsubsection{Hamming distance.}

Finally, we use the Hamming distance \citep{6772729} to assess the similarity between the subnetworks yielded by NBS. The Hamming distance measures the minimum number of substitutions required to change one subnetwork into the other.  The largest the distance, the less similar the networks are; conversely, when the distance is 0, no changes are necessary and the subnetworks are the same.

\subsection{Data and software availability}
\label{sec:data_avail}

The data that support the main findings of this study are openly available at \href{http://fcon_1000.projects.nitrc.org/indi/abide/}{\texttt{http://fcon\_1000. projects.nitrc.org/indi/abide/}}. For details about the full dataset and its procedures, see \citet{DiMartino2017, DiMartino2014}. The data that support the verification of the findings are openly available at \href{https://openneuro.org/datasets/ds000030/versions/00016}{\texttt{https://openneuro.org/datasets/ds000030/versions/00016}}, version 00016 and details about the full dataset can be found in \citet{Gorgolewski2017}. 

The preprocessing pipeline used is available at \href {https://version.aalto.fi/gitlab/BML/bramila/tree/d8d9457ad2c2b44b18710faf242c19a7a749ff4f}%
{\texttt{https://version.aalto.fi/gitlab/BML/brami la/tree/d8d9457ad 2c2b44b18710faf242c19a7a749ff4f}}, commit hash \texttt{4f1e6388d6b2e5024ef2 380d29e6526bb878242a}.

The code used to analyze the data and generate the plots in the study is available at \href {https://doi.org/10.5281/zenodo.3671882}%
{\texttt{https://doi. org/10.5281/zenodo.3671882}} (licensed under the MIT Licence).

\section{Results}
\label{sec:results}
\subsection{Smoothing affects group-level differences in functional brain networks}

\begin{figure}[!h]
\center{\includegraphics[width=\textwidth]{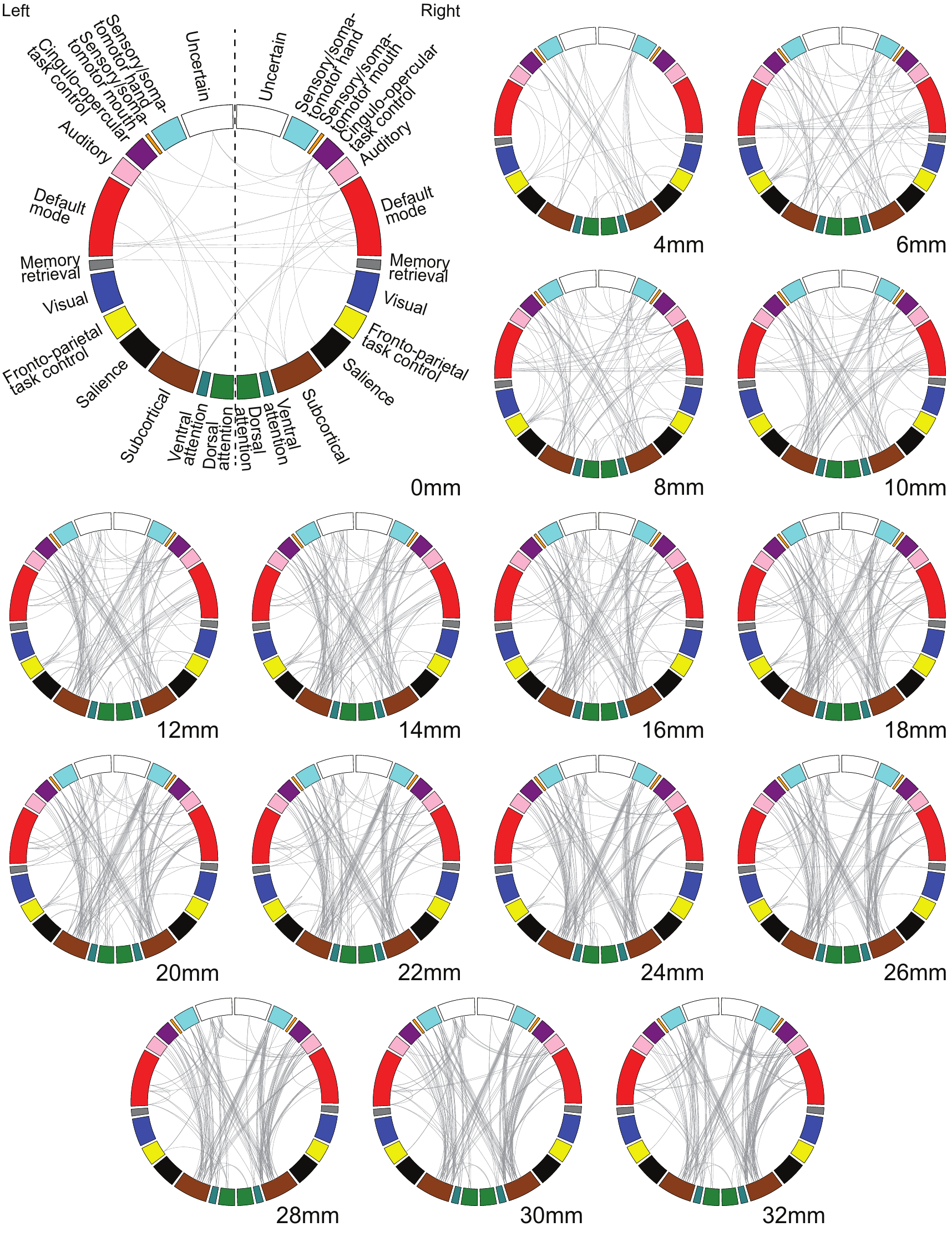}}
\caption{\label{fig:effects} Group-level differences in resting-state functional networks for different smoothing kernels. The circos plots show the between-groups connectivity differences identified by NBS. The nodes are grouped into systems following \citet{POWER2011665}, colored accordingly, and split into the left and right hemispheres. The width of the smoothing kernel changes the detected connectivity differences. The connections found at different kernel widths vary in terms of structure.}
\end{figure}

We begin our study of the effects of spatial smoothing on group-level differences by comparing the subnetworks yielded by NBS between groups (Fig. \ref{fig:effects}). We notice connectivity differences between groups for almost all systems, with no specific pattern. Yet, the detailed link structure yielded by NBS is clearly affected by the selected smoothing kernel size with links appearing and dissappearing in consecutive smoothing kernels. The number of links in the subnetwork increases with the increasing smoothing kernel FWHM across the full range of analyzed kernels.

We also compare differences in the structure of thresholded, weighted networks using permutation tests (Supplementary Fig. A2). Unlike in the NBS subnetworks, the number of significantly different links decreases as the smoothing kernel increases (see Supplementary Fig. A3 A). Moreover, the large number of significant links makes it difficult to detect possible patterns from visualizations (see Supplementary Fig. A2).

For the set of graph measures calculated on binary, thresholded networks, we find group-level differences for betweenness centrality, clustering coefficient, global efficiency, and local efficiency. However, most of those differences are found at high levels of smoothing (FWHM$>12$ mm) which are rarely used in practice. An exception is the local efficiency of the left dorsal dysgranular insula (node 173), for which we find differences at 8 mm. We did not find significant differences for the degree or the mean clustering coefficient at any smoothing level. The significant results are available in supplementary tables 8, 9, 10, and 11. The complete results (significant and not significant) are available in Zenodo (see section Data and Software availability).

\subsection{Effects of smoothing are independent of ROI size}

\begin{figure}[!h]
\center{\includegraphics[width=0.7\textwidth]{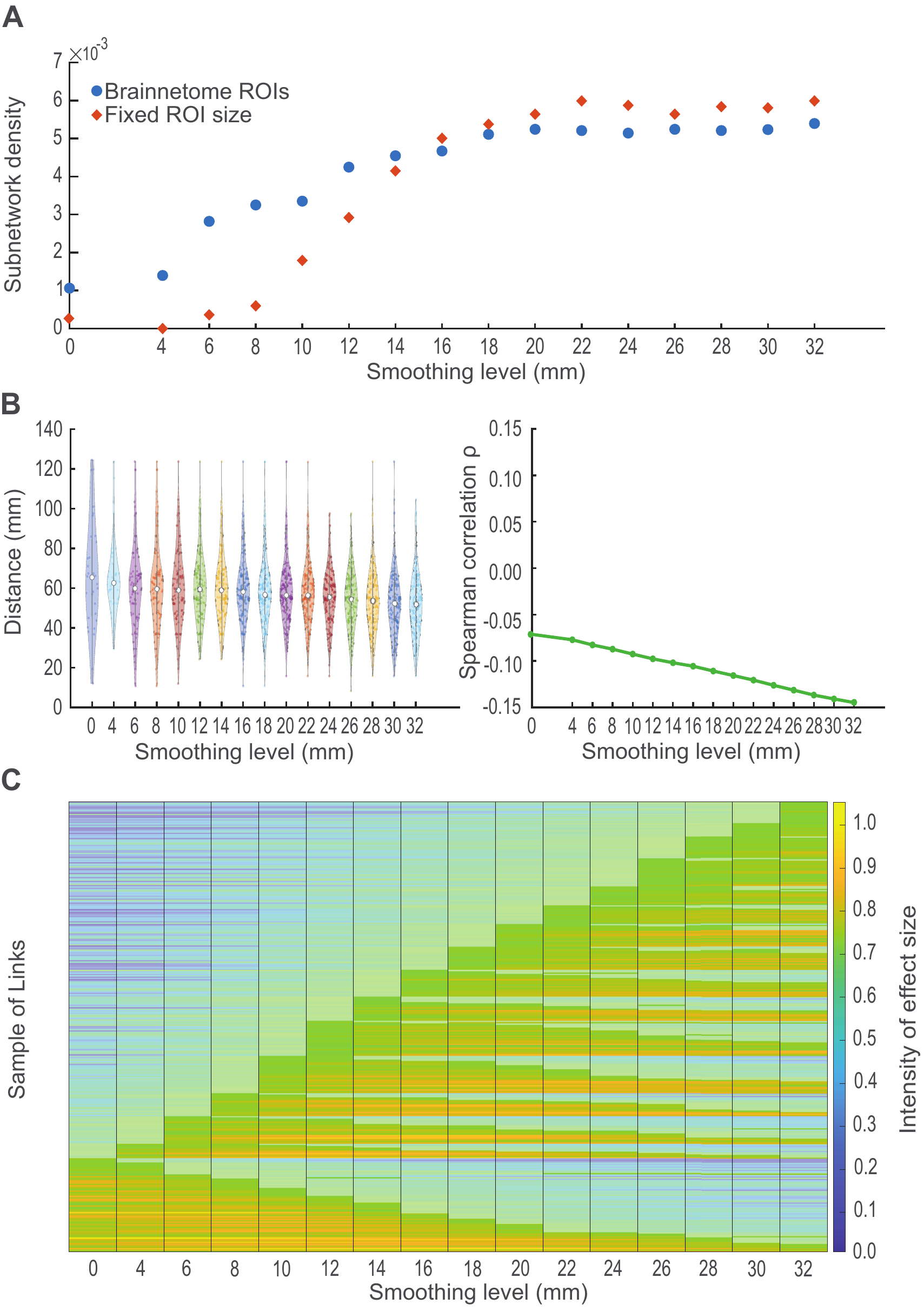}}
\caption{\label{fig:effects_investigation} Characteristics of between-group differences for each smoothing kernel. Based on the subnetworks identified by NBS, we see that: (A) The choice of smoothing kernel affects the number of links of the subnetworks and its effects are independent of ROI size. The number of links in the subnetwork increases as the smoothing level is incremented. This pattern is present even when the ROIs are artificially constructed as spheres of constant ratios. (B) Spatial smoothing alters the distance profile of detected links at commonly used kernels FWHM$\leq$12 mm with shorter links being undetected at kernels 4 mm and 12 mm. (C) Larger kernels are associated with a decrease in the detection of longer links. A decline in the value of $\rho$ with increasing FWHM highlights a stronger negative correlation between the length of the links and their F-statistic. In other words, using larger kernels decreases the chance of finding long links. (D) Effect sizes of the subnetwork links change depending on the smoothing kernel. The y-axis shows the links which NBS identified as different between groups at some kernel. The links (y-axis) are organized according to the smallest kernel in which they are detected and the number of smoothing levels in which they appear statistically significant ($\alpha$<0.05); for example, all links which are significant at 0 mm are shown at the bottom of the plot, then on top of them, we show all links which are significant at 4mm, but are not significant at 0mm. This organization follows until all smoothing levels are shown. The plot highlights those kernels at which the links are found significant. 6 links are detected at all smoothing levels. Conversely, some links are only observed when a particular kernel is used despite having a large effect at other kernels.}
\end{figure}

Next, we investigate whether the effects of smoothing are different for ROIs of different size. To examine the influence of size, we create non-overlapping spheres of constant radius centered at the Brainnetome ROI centroids. We use spherical ROIs of 7 voxels to re-compute the adjacency matrices and re-run the NBS analysis. We find that despite the equal-sized ROIs, the NBS subnetworks follow the same pattern as those identified using the Brainnetome ROIs (Fig. \ref{fig:effects_investigation} A). However, the NBS subnetwork densities are different when the fixed-sized ROIs are used instead of the full-size Brainnetome ROIs. This may be related to the fact that the fixed-size ROIs do not include all voxels of the original ROIs. This leads to differences in the time series and therefore connectivity profiles between the original and fixed-sized ROIs.

\subsection{Effects of smoothing and physical link length}

Next, we investigate the relationship between the effects of spatial smoothing and the physical distance between ROIs. We calculate the distribution of physical distances between the ROI centroids for all the links in the subnetwork at each smoothing level (Fig. \ref{fig:effects_investigation} B). We notice that some shorter links are not detected for kernels between 4, 12 and 14 mm. Conversely, a few long links are not identified for kernels 24, 26, and 32 mm. To test the influence of smoothing the data on the detection of long and short links, we calculate the Spearman correlation coefficient $\rho$ between the physical length of each link (significant and non-significant) and its F-statistic for all smoothing kernels (Fig.~\ref{fig:effects_investigation} C). We observe a negative correlation between the length of the link and its F-statistic: the longer the link, the smaller its F-statistic. We also note a decline in the values of $\rho$  as the smoothing kernel width increases, \emph{i.e.}, larger smoothing kernels are associated with smaller F-statistic for longer links, an effect we also observe in Fig.~\ref{fig:effects_investigation} B. For thresholded, weighted networks, we observe non-systematic fluctuations in the detection of long and short links (Supplementary Fig. A3 B). Nevertheless, the correlation coefficient $\rho$ reflects similar effects to those of full, weighted networks (Supplementary Fig. A3 C).

\subsection{Spatial smoothing influences effect size}

To identify the group-level differences between networks, NBS relies on a significance criterion ($\alpha<0.05$). However, the magnitude of the effect is also important \citep{Loftus1997, cohen94}. Therefore, we also explore how spatial smoothing influences the effect size of significant links detected by NBS at any level using the Cohen's d (see section Other metrics).

The effect size varies with the smoothing kernel size (Fig.~ \ref{fig:effects_investigation} D). However, this variation is not systematic: an increment in the kernel size may cause the magnitude of effect to decrease (Fig.~\ref{fig:effects_investigation} D, bottom links) or increase (Fig.~\ref{fig:effects_investigation} D, upper links). Few links have a high magnitude of effect for all smoothing kernels, and even then, these links may be discarded from the subnetworks because of the NBS significance suprathreshold (see bottom links in Fig.~\ref{fig:effects_investigation} D). In general, the variations are smooth and links are detected for a set of consecutive smoothing levels. Effects are also perceived in thresholded, weighted networks (see Supplementary Fig. A3 D). However, their effect size is smaller with sudden changes, making some links recognized at non-consecutive smoothing kernels.

\subsection{Subnetworks are more similar for some kernels}

\begin{figure}[!h]
\center{\includegraphics[width=0.6\textwidth]{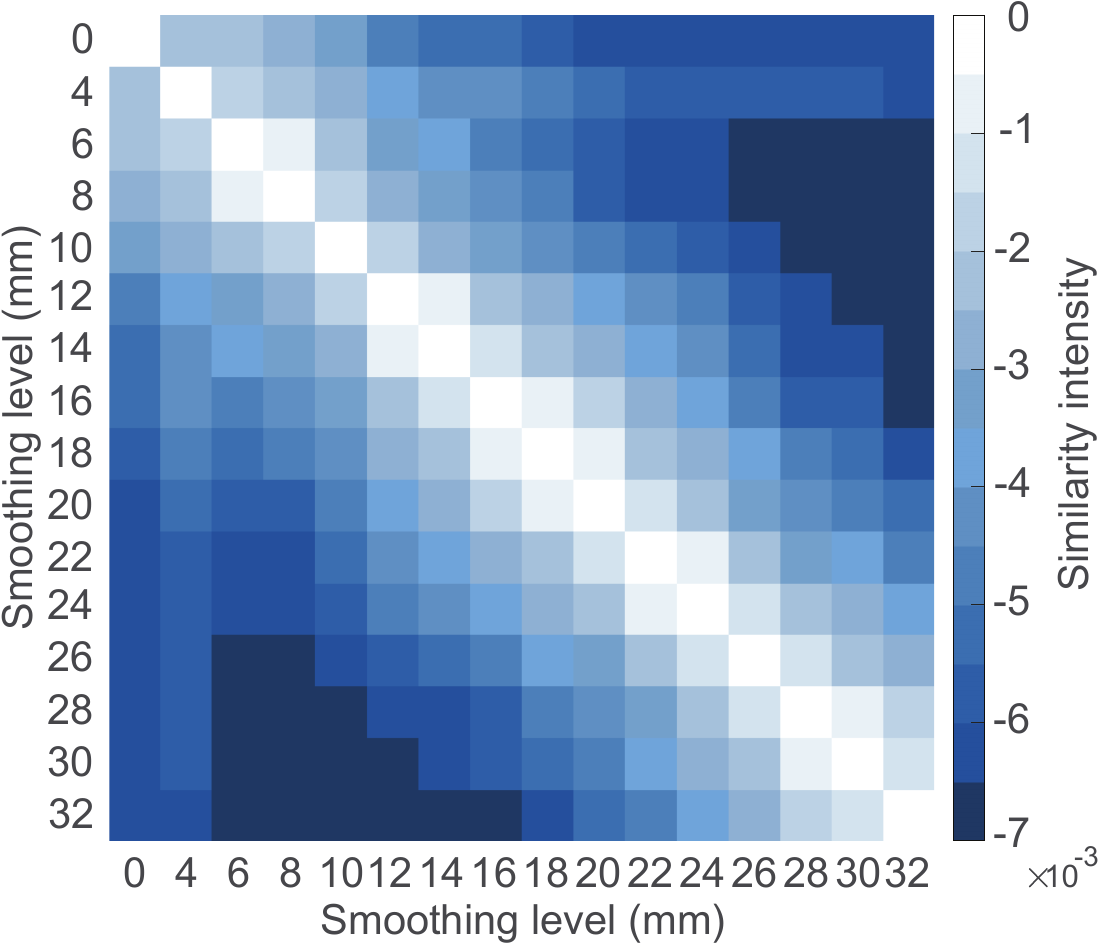}}
\caption{\label{fig:hamming} Similarity matrix between the identified subnetwork differences at each smoothing level. Brighter colors represent less differences in the structure of the subnetworks. Subnetworks pairs 6-8 mm, 12-14 mm, 16-18 mm, 18-20 mm, 22-24 mm, and 28-30 mm are more similar than the rest of kernels.}
\end{figure}

Finally, we examine how similar the subnetworks are at different smoothing levels. To inspect this similarity, we compute the Hamming distance between the subnetworks (see section Other metrics) (Fig.~\ref{fig:hamming}). As one might expect, the pairs of subnetworks obtained for consecutive kernels (6-8 mm, 12-14 mm, 16-18 mm, 18-20 mm, 22-24 mm, and 28-30 mm) exhibit greater similarity, while the subnetwork pairs that show largest differences are obtained with kernels far away from each other (e.g. 6-26 mm, 8-26 mm, 10-28 mm, 12-30 mm, 14-32mm). For thresholded, weighted networks, the least similar subnetwork pairs are obtained with small kernels, particularly 0-8 mm, (Supplementary Fig. A4). This is surprising, since the least similar subnetworks are found for the commonly used kernels, and particularly between commonly used kernels and unsmoothed data. Therefore, even a small change in kernel size may lead to drastic changes in the observed differences between groups.

\subsection{Generalizability of the results}

To demonstrate the generality of our results, we repeated the analysis after changing several aspects of the pipeline: i) the NBS suprathreshold value, ii) the parcellation scheme and number of ROIs: two Craddock parcellations (\citet{Craddock2012}, see Methods section) instead of Brainnetome, iii) the network density used to study weighted and binary thresholded networks, and iv) the spatial smoothing method: adaptive smoothing (\citet{AFNI}, see Methods section) instead of the non-adaptive Gaussian smoothing. Besides, we demonstrated that the results generalize for a second, independent dataset, the UCLA dataset (see Methods section).

Our main results are robust against changes in these parameters: the level of spatial smoothness affects group-level differences in brain network structure in nontrivial and unexpected ways. However, parameter choices, in particular the values of the NBS suprathreshold and network density, affect the details of observed between-group differences. The combined effects of these parameters together with spatial smoothing are even harder to predict than those of smoothing alone. For further details, see the section \emph{Generality of the results} in the Supplementary Material and Supplementary Figures A5--A44.

\section{Discussion}

In this work, we investigated the effects of spatial smoothing on group-level differences in functional brain networks. Resting-state fMRI data from clinical populations (ASD and bipolar disorder) and matched controls (TC) were spatially smoothed with Gaussian filters with increasing smoothing kernel width. Functional networks for each subject were estimated using Pearson’s correlation coefficients and group differences were computed for the full correlation matrix using the Network-based Statistic Toolbox (NBS). We also reproduced the findings for weighted, thresholded functional brain networks. Our results establish three findings: (i) the choice of spatial smoothing kernel affects group-level differences in resting-state functional brain networks, (ii) some links are significantly different between the two groups at all kernel sizes, while most links are different only at specific smoothing kernels, (iii) the effects are not explained by the spatial properties of the parcellation used and are in general non-systematic and difficult to predict. Furthermore, the graph-theoretical properties of the functional nodes did not produce systematic group differences in the ABIDE dataset. However, differences in the degree of nodes right medial area 8 and left caudal temporal thalamus were found for the UCLA data.

\subsection{Group differences increase with kernel width for commonly used kernels}
Spatial smoothing is a crucial preprocessing step for GLM analysis and registration \citep{Mikl2018}, but is it vital for connectivity analysis? Here we showed that the number of significantly different links increases with kernel sizes (Fig. \ref{fig:effects_investigation} A, blue). It is expected that a moderate level of spatial smoothing can increase the SNR of fMRI time series as it can control some undesired effects like head motion and thermal noise \citep{Power2017, Constable}. Consequently, spatial smoothing should also enhance the detectability of group differences in connectivity. However, it was surprising to identify a large number of links that were significantly different above the commonly used smoothing kernels of 4--12 mm. As large kernels mix signals from distant brain regions, increasing the similarity between all node time series \citep{doi:10.1111/ejn.13717}, asymptotically reaching the global signal with all nodes having the same identical temporal dynamics, one might expect seeing group differences to disappear at large kernels. However, this still remains to be confirmed, given there were differences even at FWHM=32 mm (Fig. \ref{fig:effects}). In practice, these large kernels are never used.

A possible explanation of fairly large kernels still producing group differences is that fMRI signals processed with kernels larger than 12 mm might reflect mesoscopic fluctuations, comparable to the average level of activity of functional modules. In contrast, data with no smoothing emphasizes localized voxel activity. It has been shown in ASD that modular differences play a more important role than single link differences \citep{doi:10.1002/hbm.23084}, with similar modules also detected in our analysis (subcortical areas as well as parts of the DMN).

\subsection{Which links are significantly different?}

When examining which links were significantly different between the two groups, we found 6 links were consistently different at all smoothing levels and 37 links were consistent at 12 smoothing levels, which represents 75\% of the levels; these 6 most stable links are between the subcortical regions (thalamus and basal ganglia), superior frontal gyrus, precentral gyrus, paracentral lobule, and postcentral gyrus; and fusiform gyrus and postcentral gyrus. At the most common FWHM values, most of the significant connections are diverse, indicating global connectivity changes, which as expressed by \citet{Hull2016} suggests that individuals' uniqueness of ASD hinders finding a predominant trend. In accordance with many studies (see \citep{Hull2016, Maximo2014} for a full review), we find differences in connectivity of regions related to the DMN, ventral and dorsal attention, and motor areas. However, we find few results between DMN regions. This discrepancy could be explained by methodological issues like patient cohort, network construction method, or statistical analysis, not to mention the hetereogenous nature of ASD \citep{Ha2015a, Maximo2014,VASA2016245}. 
Moreover, we should also keep in mind the kernel width when comparing the results, since the subnetwork composition changes accordingly. 
When looking at graph-theoretical properties of nodes computed on thresholded networks, no conclusion is reached for the effects on degree and mean clustering coefficient, since we did not find differences between groups at any smoothing levels, despite previous findings in the literature \citep{Maximo2014}.

Similarly to the ASD study, 31 links are consistently different at 75\% of the smoothing levels for the case of bipolars vs.~controls, 8 of which are found across the 16 levels. However, the larger amount of significant links detected makes it difficult to find specific patterns. Despite finding links connecting most of the systems, we notice numerous connections involving the cingulo-opercular, auditory, and sensorimotor hand areas. Notably, we find differences in the connectivity of dorsal attention areas, but no connections in the ventral attention areas. Contrary to previous findings, we find differences in the DMN and salience networks (see \cite{Syan2018} for full review). 

\subsection{Results generalize well}

When comparing group differences between populations, there are many choices of parameters that may affect the results and their generality. To demonstrate the validity of our conclusions, we conducted several analyses addressing different parcellations, NBS suprathresholds, and network densities for our methods.

Results for different parcellation schemes are similar, excluding some variation in the effects of spatial smoothing on links of different physical length (Supplementary Fig. A11). Moreover, the results remained unaffected when controlling for spatial properties of the parcellations, in particular using equal volumes for all ROIs or taking the distance between ROIs into account (Supplementary Fig. A8). The results are also fairly robust for a range of numbers of ROIs. The lower limit of this range is determined by the mesoscopic, or voxel-level, characteristics of the data one wants to highlight. When the number of ROIs gets low enough, their functional interpretation may change. For example, ROIs of the Craddock100 parcellation are most probably rather larger brain systems than functionally homogeneous areas specialized on certain tasks like the ROIs of higher-resolution parcellations. Therefore, it is not surprising that the results obtained with Craddock100 were somewhat different from those obtained with other parcellations.

As expected, NBS suprathreshold and network density impact the amount of significant links. For weighted, thresholded matrices, the density threshold might remove a link for one subject, while retaining its only slightly higher weight for another. The combined effects of smoothing and thresholding are indeed irregular and hard to predict. Likewise, the possible non-linearities in the NBS change the results obtained for different smoothing kernels, as the list of links that comprise the subnetwork is influenced by the suprathreshold level. The effect of the suprathreshold on the NBS outcome is not unique to the present study. Indeed, the developers of NBS recommend tuning the suprathreshold for the data at hand or investigating a range of suprathreshold values \citep{ZALESKY20101197}. Moreover, spatial smoothing influences the effect size of the individual links irregularly. For NBS, changes in the effect size of a link may make it not compliant with the suprathreshold level, discarding it from the subnetwork.

Using data from different sources is not uncommon in brain network analysis of clinical populations. Here, we use ABIDE data from different scan centers to make our results more reproducible. However, the highly imbalanced number of subjects from the sites could pose a problem, despite regressing possible site-effect confounds in our analysis of the ABIDE data. To further address the multisite effects, we have replicated our study in another dataset whose images were taken at the same site and in the same scanner (UCLA). Remarkably, the results obtained from this second dataset are close to those obtained from the ABIDE data, indicating that the observed effects do not relate to site effects and are not exclusive for one dataset.

We have seen the unpredictability of the effects of spatial smoothing, which is an optional but commonly used fMRI preprocessing step. But what are the effects of the spatial smoothness introduced by other preprocessing steps like interpolation? And what is its role in data sets like ABIDE, whose images come from different scan sites and may have different amounts of intrinsic smoothness? In the case of ABIDE, this may pose problems, as the images come from different sites. In general, if we know the intrinsic smoothness for each image, i.e. the smoothness related to the acquisition and processing steps like interpolation, we can apply different kernel sizes accordingly. We also explored this adaptive smoothing, with the finding that the observed effects remain when all images have the same final smoothness. At small smoothing kernels (FWHM$\leq$6 mm), adaptive smoothing seems to yield more consistent NBS subnetworks (Supplementary Fig. A31) compared to the non-adaptative method.

\subsection{Limitations and future directions}
Some limitations should be taken into consideration when trying to generalize the findings. First, the lack of ground truth for the tested comparisons makes it hard to recommend the kernel width that best identifies actual group differences. It would be interesting to run a similar analysis using test-retest data for the same subjects, which could help to elucidate which differences are more likely due to true contrasts between groups. Second, the effects of spatial smoothing may vary depending on how the ROI time series have been defined. Here, we use the Pearson correlation as the definition, but other methods like the first principal component could be employed. How this definition alters the current results could be explored in future works.

Finally, in many parcellations, the ROI size varies. Smoothing mixes signals across ROI boundaries, and small ROIs are more prone to this mixing than larger ones. Thus, we tested whether the size of the ROIs had some impact on the results. By having non-overlapping, fixed-size ROIs, we did not take into account those voxels that are close to the ROIs boundaries and that are most likely to have signal from other ROIs. In the future, other methods like binary-eroding the ROIs could be tested as well.

\subsection{Conclusions}
To smooth or not to smooth? The decision might still depend on the goals of the researcher. While it has been recommended to avoid smoothing in multivoxel pattern analysis \citep{KAMITANI20101949,Mahmoudi2012}, other authors have suggested to adopt a kernel width of three times the voxel size for GLM analysis \citep{lindquist2008}, or twice the voxel size for task-based inter-subject correlation analysis \citep{Pajula2018}. Here, we have seen that although spatial smoothing the data affects the result similarly for different parcellations and datasets, there is no unique level that will fit every case.  

Functional connectivity studies often rely on brain parcellations to reduce the spatial dimensionality of the data, which in practice apply a level of smoothing proportional to the size of the ROIs. Due to the lack of ground truth in clinical resting-state fMRI connectivity patterns, it is difficult to say whether spatial smoothing causes non-existing disparities to be viewed as significant findings, or whether it improves data quality so that true differences between the networks are revealed. Likewise, it is challenging to say whether we should aim for those kernel values (or smoothing strategies) that yield more consistent results across smoothing kernel FWHMs, or whether we should exploit this trait to improve group distinction \citep{borschart}. In general, the effects of spatial smoothness are complex and difficult to predict. Hence, spatial smoothing should be considered carefully, as it alters network differences when comparing functional brain networks of different groups.

\section*{Supporting information}
Supplementary data associated with this article can be found in the \textit{supplementary files} folder at \href {https://doi.org/10.5281/zenodo.3671882}%
{\texttt{https://doi.org/10.5281/zenodo.3671882}}, as the two separated files: Supplementary tables and Supplementary information.

\section*{Acknowledgements}
We acknowledge the computational resources provided by the Aalto Science-IT project.

\section{Funding information}
O.K. has been supported by The Osk. Huttunen Foundation and The Emil Aaltonen Foundation.

\bibliographystyle{apalike}
\bibliography{bibliography}

\end{document}